\documentstyle[prd,aps,eqsecnum,preprint,tighten,floats,psfig]{revtex}

\begin{document}

\draft

\preprint{\vbox{\hfill UMN-D-01-10 \\
          \vbox{\hfill OHSTPY-HEP-T-01-031} \\
          \vbox{\hfill hep-th/0112151}
          \vbox{\vskip0.3in}
          }}

\title{Simulation of Dimensionally Reduced SYM--Chern--Simons Theory}

\author{J. R. Hiller}
\address{Department of Physics, 
University of Minnesota Duluth, 
Duluth, MN 55812}

\author{S.S. Pinsky and U. Trittmann}
\address{Department of Physics, 
The Ohio State University, 
Columbus, OH 43210}

\date{\today}

\maketitle

\begin{abstract}
A supersymmetric formulation of a three-dimensional SYM--Chern--Simons 
theory using light-cone quantization is presented, and the supercharges 
are calculated in light-cone gauge.  The theory is dimensionally reduced 
by requiring all fields to be independent of the transverse dimension. 
The result is a non-trivial two-dimensional supersymmetric theory with an 
adjoint scalar and an adjoint fermion. We perform a 
numerical simulation of this SYM--Chern--Simons theory in 
1+1 dimensions using SDLCQ (Supersymmetric Discrete Light-Cone Quantization). 
We find that the character of the bound states of this theory is very
different from previously considered two-dimensional 
supersymmetric gauge theories. The low-energy bound states of this theory are
very ``QCD-like.'' The wave functions of some of the low mass states have a 
striking valence structure. We present the valence and sea parton structure 
functions of these states. 
In addition, we identify BPS-like states which are almost independent of
the coupling. Their masses are proportional to their parton number 
in the large-coupling limit.

\end{abstract}

\narrowtext

\section{Introduction}

Chern--Simons (CS) theories are certainly some of the most interesting 
field theories in 2+1 dimensions.  Among the interesting phenomena one 
sees in these theories are: the quantum hall effect, Landau levels, 
non-trivial topological structures, vortices, and anyons. For a review of
these phenomena see \cite{dunne}. Some of these (2+1)-dimensional
phenomena have been observed experimentally in condensed matter
systems.  CS theories are also often described as topological theories
\cite{poly1} since for most gauge groups the CS coupling must obey a 
quantization condition for the theory to remain gauge invariant.  Within
this rich literature on CS theories there is also considerable work on
SYM--CS theories. These theories have their own
remarkable properties. It has been shown that there is a finite anomaly
that shifts the CS coupling \cite{lee}, and it has been conjectured by
Witten \cite{witten2} that this theory spontaneously breaks supersymmetry
for some values of the CS coupling. There are other related reasons for
interest in CS theories.  Witten \cite{witten} has conjectured that string
field theory is essentially a non-commutative CS theory.  Recently this
led to a conjecture by Susskind \cite{susskind} that relates string
theory to the fractional quantum Hall effect.

Since these theories are interesting from so many different points of 
view, it is certainly useful to  numerically simulate them. The method  we
will use is SDLCQ (Supersymmetric Discrete Light-Cone Quantization).  This
is a numerical method that can be used to solve any theory with enough
supersymmetry to be finite.  The central point of this method is that
using DLCQ we can construct a finite dimensional representation of the
superalgebra \cite{sakai95}. From this representation of the superalgebra, 
we construct a finite-dimensional Hamiltonian which we diagonalize 
numerically. We repeat the process for larger and larger representations
and extrapolate the solution to the continuum.  We have already solved
standard (2+1)-dimensional supersymmetric Yang--Mills (SYM)  theories by
this method \cite{hpt2001,hpt2001b}, so it is clear that SYM-CS
theories are within our reach. 

In this paper we will start with SYM--CS theory in 2+1 dimensions and
dimensionally reduce it to two dimensions by requiring all of the fields 
to be independent of the transverse coordinate. This is an interesting 
(1+1)-dimensional supersymmetric theory in its own right. It is a good 
starting point for our numerical simulations because we can solve this
problem using two completely independent codes, one a {\sc Mathematica}
code and the other a C++ code. When we move on to 2+1 dimensions in
future work, the (1+1)-dimensional results will be related to the
(2+1)-dimensional results in a nontrivial way. We found  this to be the
case in our previous work on (2+1)-dimensional SYM theories,  and it will
most probably be the case here as well. This will provide us an important check
on our code for solving the (2+1)-dimensional problem. It is important to
develop this chain of numerical checks, since there are no analytical or
other numerical solutions to check our results against.

Many of the most interesting aspects of CS theory will be lost by this 
reduction, most notably the quantization of the CS coupling. However,  one
particularly interesting property that will be preserved is the fact  that
the CS term simulates a mass for the theory. It is well known that
supersymmetric abelian CS theory is simply the theory of a free massive
fermion and a free massive boson. In the non-abelian theory additional
interactions are introduced, but we expect to also see this  mass effect.
This is particularly interesting here because dimensionally  reduced
${\cal N}=1$ SYM is a  very stringy theory.  The low-mass states are
dominated by Fock states with many constituents, and as the size of the
superalgebraic representation is increased, states with lower masses and
more constituents appear \cite{hpt2001,hpt2001b,alpt98,Lunin:2001im,%
Pinsky:2000rn,Hiller:2000nf,Haney:2000tk,Lunin:1999ib,Antonuccio:1999zu,%
Antonuccio:1998mq,Antonuccio:1998tm,Antonuccio:1998jg,Antonuccio:1998kz}. 
The connection between string theory and supersymmetric gauge theory leads 
one to expect this type of behavior; however, these gauge theories are not
very QCD-like.  Ultimately one might like to make a connection with the
low-mass spectrum observed in nature. 

We expect to find states in SYM--CS theory that are much more
QCD-like, because of the effective mass for the constituents introduced
by the CS term.  We will see that some of the low-mass states will be 
dominated by valence-like Fock states that have only a few constituents.
As we go to larger and larger representation of the superalgebra, we will
resolve these states better and better, and any new states that appear will
be heavier.  We will also find highly mixed states without a valence
structure.  Finally, we find that at strong coupling the low-energy 
spectrum is dominated by states that are a reflection of the BPS states 
of the SYM theory in 1+1 dimensions \cite{Antonuccio:1998kz} and are 
therefore independent of the coupling $g$.

>From the wave functions of the bound states we will be able to find the
structure functions of both the valence and sea partons. Some of the highly
mixed states will have a double-humped structure function. Some of these
structure functions are similar to those conjectured in various
phenomenological calculations; they are the result of the solution of a
nontrivial gauge theory. 

In Sec.~\ref{sec:SuperCS} we give a general discussion of supersymmetric
light-cone-quantized SYM--CS theory in light-cone gauge. We then give the 
dimensionally reduced discrete formulation of the supercharges and discuss 
the other symmetries of the theory. In Sec.~\ref{sec:Results} we discuss 
our numerical method and some of the new wrinkles that appear in CS
theory.  We also present and discuss in this section the spectrum of bound
states  as well as a variety of properties of these states including their
structure functions. We will see that it is a very QCD-like theory as 
opposed to pure SYM theory, which is very stringy. Finally, in 
Sec.~\ref{sec:Summary}, we summarize the results and discuss the prospects 
and challenges for calculating in the full (2+1)-dimensional theory.

\section{Supersymmetric Chern--Simons theory} \label{sec:SuperCS}

We will consider ${\cal N}=1$ supersymmetric CS theory in 2+1 dimensions as 
the starting point of our discussion. While we will reduce this theory 
to a (1+1)-dimensional theory for numerical simulation here, we eventually 
plan to simulate the full theory, and it is therefore useful to present a 
detailed light-cone formulation in light-cone gauge.  The Lagrangian 
of this theory is 
\begin{equation}
{\cal L}={\rm Tr}(-\frac{1}{4}{\cal L}_{\rm YM}+i{\cal L}_{\rm F}
+\frac{\kappa}{2}{\cal L}_{\rm CS}),\label{Lagrangian}
\end{equation}
where $\kappa$ is the CS coupling and
\begin{eqnarray}
{\cal L}_{\rm YM}&=&F_{\mu\nu}F^{\mu\nu}\,, \\
{\cal L}_{\rm F}&=&\bar{\Psi}\gamma_{\mu}D^{\mu}\Psi\,, \\
{\cal L}_{\rm CS}&=&\epsilon^{\mu\nu\lambda}\left(A_{\mu}
\partial_{\nu}A_{\lambda}+\frac{2i}{3}gA_\mu A_\nu A_\lambda \right)
+2\bar{\Psi}\Psi\,. \label{eq:CSLagrangian}
\end{eqnarray}
The two components of the spinor $\Psi=2^{-1/4}({\psi \atop \chi})$ 
are in the adjoint representation of $U(N_c)$ or $SU(N_c)$.
We will work in the large-$N_c$ limit. 
The field strength and the covariant derivative are
\begin{equation}
F_{\mu\nu}=\partial_{\mu}A_{\nu}-\partial_{\nu}A_{\mu}
              +ig[A_{\mu},A_{\nu}]\,, \quad \quad 
D_{\mu}=\partial_{\mu}+ig[A_{\mu},\quad]\,.
\end{equation} 
The supersymmetric variations of the fields are
\begin{eqnarray}
\delta A_\mu&=&i\bar{\epsilon}\gamma_{\mu}\Psi\,,  \\
\delta\Psi&=&\frac{1}{4}i\epsilon^{\mu\nu\lambda}\gamma_\lambda F_{\mu\nu}
=\frac{1}{4}\Gamma^{\mu\nu}\epsilon F_{\mu\nu}\,,
\end{eqnarray}
where\footnote{This choice of representation for the Dirac matrices
is different from that of Ref.~\cite{dunne} by the interchange of
$\gamma^1$ and $\gamma^2$ but is more natural for light-cone quantization.
In this representation the spinor term of the CS
Lagrangian enters Eq.~(\ref{eq:CSLagrangian}) with a plus sign.}
\begin{equation}
\gamma^0=\sigma_2, \quad \gamma^1=i\sigma_1,\quad
\gamma^2=i\sigma_3, \quad \Gamma^{\mu\nu}\equiv
\frac{1}{2}\{\gamma^\mu,\gamma^\nu\}= 
          i \epsilon^{\mu\nu\lambda} \gamma_{\lambda}\,.
\end{equation}
This leads to the supercurrent $Q^{(\mu)}$ in the usual manner via  
\begin{equation}
\delta{\cal L}=\bar{\epsilon}\partial_{\mu}Q^{(\mu)}.
\label{LQ}
\end{equation}
Light-cone coordinates in 2+1 dimensions are $(x^+,x^-,x^\perp)$ where 
$x^+=x_-$ is the light-cone time and $x^\perp=-x_\perp$. The totally 
anti-symmetric tensor is defined by $\epsilon^{+-2}=-1$.
The variations of the three parts of the Lagrangian in Eq.~(\ref{Lagrangian}) determine the (`chiral') components $Q^\pm$ of the supercharge
via Eq.~(\ref{LQ}) to be
\begin{equation}
\int d^2x Q^{(+)}=\left( {Q^+ \atop Q^-}\right)
   = \frac{i}{2}\int d^2x\,
\Gamma^{\alpha\beta}\gamma^{+}\Psi F_{\alpha\beta}\,.
\end{equation}
Explicitly they are 
\begin{eqnarray} \label{supercharges}
Q^-&=&-i 2^{3/4}\int d^2x\,
\psi\left(\partial^+ A^- -\partial^-A^+ + ig[A^+,A^-]\right)\,,
\nonumber \\
Q^+&=&-i 2^{5/4}\int d^2x\,
\psi\left(\partial^+ A^2 -\partial^2 A^+ + ig[A^+,A^2]\right)\,.
\end{eqnarray}
One can convince oneself by calculating the energy-momentum tensor
$T^{\mu\nu}$ that the supercharge fulfills the supersymmetry algebra
\begin{equation}
\{Q^\pm,Q^\pm\}=2\sqrt2 P^\pm\,, \qquad 
\{Q^+,Q^-\}=-4P^\perp\,.
\end{equation}

In order to express the supercharge in terms of the physical degrees of
freedom, we have to use equations of motion, some of which are
constraint equations. The equations of motion for the gauge fields are
\begin{equation}
D_\nu F^{\nu\alpha}=-J^\alpha\,,
\end{equation}
where 
\begin{equation}
J^\alpha=\frac{\kappa}{2}\epsilon^{\alpha\nu\lambda}
F_{\nu\lambda}+2g\bar{\Psi}\gamma^\beta\Psi\,.
\end{equation}
For $\alpha=+$ this is a constraint for $A^-$,
\begin{equation}
D_-A^-=-(D_2-\kappa)A^2-\frac{1}{D_-}(D_2-\kappa)\partial_2 A^+
             +2g\frac{1}{D_-}\bar{\Psi}\gamma^+\Psi\,.
\end{equation}
In light-cone gauge, $A^+=0$, this reduces to
\begin{equation}
D_-A^-=\frac{1}{D_-}[(\kappa-D_2)D_-A^2+2g\bar{\Psi}\gamma^+\Psi]\,.
\end{equation}
The equation of motion for the fermion is
\begin{equation}
\gamma^\mu D_\mu \Psi=-i\kappa\Psi\,.
\end{equation}
Expressing everything in terms of $\psi$ and $\chi$ leads to the 
equations of motion
\begin{eqnarray}
\sqrt{2}D_+\psi&=&(D_2+\kappa)\chi\,,  \\
\sqrt{2}D_-\chi&=&(D_2-\kappa)\psi\,,
\end{eqnarray}
the second of which is a constraint equation.  The constraint equations
are used to eliminate $\chi$ and $A^-$.

We now reduce the theory dimensionally 
to two dimensions by setting $\phi=A_2$ and
$\partial_2 \rightarrow 0$ for all fields. This yields,
from Eq.~(\ref{supercharges}) and the constraints,
\begin{equation}
Q^-=2^{3/4}g\int dx^- \left(i[\phi,\partial_-\phi]
         +2\psi\psi-\frac{\kappa}{g}
                  \partial_-\phi\right)\frac{1}{\partial_-}\psi\,.
\end{equation}
The mode expansions in two dimensions are
\begin{eqnarray}
\phi_{ij}(0,x^-) &=& 
\frac{1}{\sqrt{2\pi}}
\int_0^\infty
         \frac{dk^+}{\sqrt{2k^+}}\left[
         a_{ij}(k^+)e^{-{\rm i}k^+x^-}+
         a^\dagger_{ji}(k^+)e^{{\rm i}k^+x^-}\right]\,,
\nonumber\\
\psi_{ij}(0,x^-) &=&\frac{1}{2\sqrt{\pi}}\int_0^\infty
         dk^+\left[b_{ij}(k^+)e^{-{\rm i}k^+x^-}+
         b^\dagger_{ji}(k^+)e^{{\rm i}k^+x^-}\right]\,.
\end{eqnarray}
To discretize the theory we impose periodic boundary conditions on the
boson and fermion fields alike, and obtain an expansion of the fields
with discrete momentum modes.
Thus the discrete version of the CS part of the supercharge is
\begin{equation} \label{qcs}
Q^-_{CS}=\left(\frac{ig2^{-1/4}\sqrt{L}}{\pi}\right)
(-ih)\sum_{n}\frac{1}{\sqrt{n}}
\left(A^{\dagger}(n)B(n)+B^{\dagger}(n)A(n)\right),
\end{equation}
where $h\equiv \sqrt{\pi}\kappa/g$ is a rescaled CS coupling 
and $A$ and $B$ are rescaled discrete field operators
\begin{equation}
A(n)\equiv \sqrt{\frac{\pi}{L}}a_{ij}(n\pi/L)\,,  \quad\quad
B(n)\equiv \sqrt{\frac{\pi}{L}}b_{ij}(n\pi/L)\,.
\end{equation}
The ordinary supersymmetric part of the supercharge is listed 
elsewhere \cite{hpt2001}.  It is important to note that the supercharge 
for ${\cal N}=1$ SYM in 2+1 dimensions has a contribution of the form
\begin{equation}
Q^-_{\perp}=\left(\frac{ig2^{-1/4}\sqrt{L}}{\pi}\right)
                \sum_{n,n_\perp}\frac{n_\perp}{\sqrt{n}}
                     \left(A^{\dagger}(n,n_\perp)B(n,n_\perp)
                         -B^{\dagger}(n,n_\perp)A(n,n_\perp)\right)\,.
\end{equation}
Of course, the light-cone energy is $(k^2_\perp + m^2)/k^+$,
so $k_\perp$ behaves like a mass, and here we see that $h$ appears in a 
very similar way to $k_\perp$ and therefore behaves in many ways like a mass.

When comparing the two contributions to the supercharge, we see that we
have a relative $i$ between them. Thus the usual eigenvalue problem 
\begin{equation}
2P^+P^-|\varphi\rangle=\sqrt{2}P^+(Q^-)^2|\varphi\rangle=
\sqrt{2}P^+(Q^-_{\rm SYM}+Q^-_{\rm CS})^2|\varphi\rangle=M^2_n|\varphi\rangle
\label{EVP}
\end{equation}
has to be solved by using fully complex methods. 

We retain\footnote{We note that in three dimensions the CS term breaks 
transverse parity.}
the $S$-symmetry, which is associated with the orientation of the 
large-$N_c$ string of partons in a state \cite{kutasov93}. In a 
(1+1)-dimensional model this orientation parity is usually referred as a $Z_2$
symmetry, and we will follow that here.  It gives a sign when the color
indices are permuted
\begin{equation}\label{Z2}
Z_2 : a_{ij}(k)\rightarrow -a_{ji}(k)\,, \qquad
      b_{ij}(k)\rightarrow -b_{ji}(k)\,.
\end{equation}
We will use this symmetry to reduce the Hamiltonian matrix size and
hence the numerical effort. All of our states will be labeled by the
$Z_2$ sector in which they appear. We will not attempt to label the states 
by their normal parity; in the light-cone this is only an approximate 
symmetry. Such a labeling could be done in an approximate way,
as was shown by Hornbostel \cite{horn}, and might be useful for 
comparison purposes if at some point there are results from 
lattice simulations of the present theory. 

\section{Numerical results}  \label{sec:Results}

We convert the mass eigenvalue problem $2P^+P^-|M\rangle = M^2 |M\rangle$ 
to a matrix eigenvalue problem by introducing
a basis where $P^+$ is diagonal.
In SDLCQ this is done by first discretizing the supercharge $Q^-$
and then constructing $P^-$ from the square of the supercharge:
$P^- = (Q^-)^2/\sqrt{2}$.  To discretize the supercharge, we introduce
discrete longitudinal momenta $k^+$ as fractions $nP^+/K$ of the
total longitudinal momentum $P^+$, where $K$ is an integer that
determines the resolution of the discretization and is known
in DLCQ as the harmonic resolution~\cite{pb85}.
Because light-cone longitudinal momenta are always positive,
$K$ and each $n$ are positive integers; the number of constituents
is then bounded by $K$.  The integrals in $Q^-$ are approximated by
a trapezoidal form.  The continuum limit 
is then recovered by taking the limit $K \rightarrow \infty$. 

In constructing the discrete approximation we drop the
longitudinal zero-momentum mode.  For some discussion of
dynamical and constrained zero modes, see the review~\cite{bpp98} 
and previous work~\cite{alpt98}.
Inclusion of these modes would be ideal, but the techniques
required to include them in a numerical calculation
have proved to be difficult to develop, particularly because
of nonlinearities.   For DLCQ calculations that can be 
compared with exact solutions, the exclusion of
zero modes does not affect the massive spectrum~\cite{bpp98}.
In scalar theories it has been known for some time that 
constrained zero modes can give rise to dynamical symmetry 
breaking~\cite{bpp98}, and work continues on the role of zero modes and near 
zero modes in these theories~\cite{thorn}.  

To obtain the spectrum of the CS theory we solve the 
complex eigenvalue problem, Eq.~(\ref{EVP}).
For the numerical evaluation we can exploit the structure of the supercharge 
\begin{equation}
Q^-=\left(\begin{array}{cc} 0 &A+iB\\ A^T-i B^T&0
\end{array}\right),
\end{equation}
where $A$ and $B$ are real matrices. The Hamiltonian has thus an easy 
decomposition into a real and imaginary part in the bosonic sector
\begin{equation}
P^-_{\rm boson}=AA^T+BB^T+i\left(BA^T-AB^T\right)\,.
\end{equation}
and the fermionic sector
\begin{equation}
P^-_{\rm fermion}=A^T A+B^T B+i\left(A^T B-B^T A\right)\,,
\end{equation}

Our earliest SDLCQ calculations were done using a code written in
{\sc Mathematica} and performed on a PC. This code was then rewritten in 
C++ and substantially revised.  It runs on a Linux workstation 
with 2 GB of RAM and can handle 
as many as 2,000,000 Fock states.  The present calculation was done in 
both codes as a check. We limit the calculation to resolution $K=9$ 
because it seems sufficient here and because the present C++ code,
which was primarily written for (2+1)-dimensional models, would require 
some non-trivial modifications to run at higher longitudinal resolutions. 

\begin{figure}
\begin{tabular}{cc}
\psfig{figure=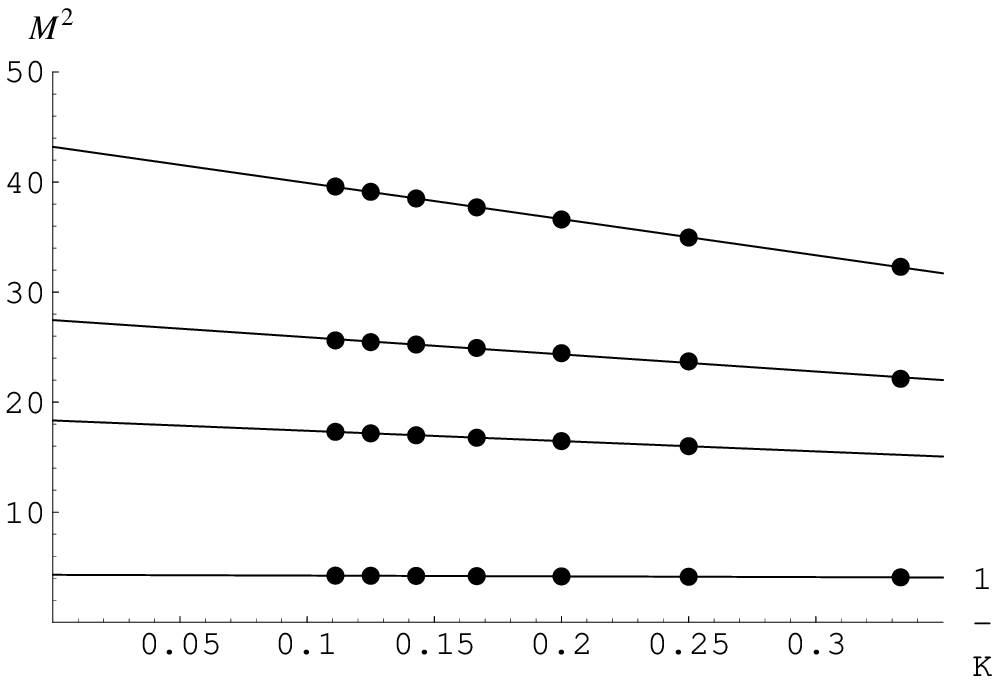,width=7.5cm,angle=0} &
\psfig{figure=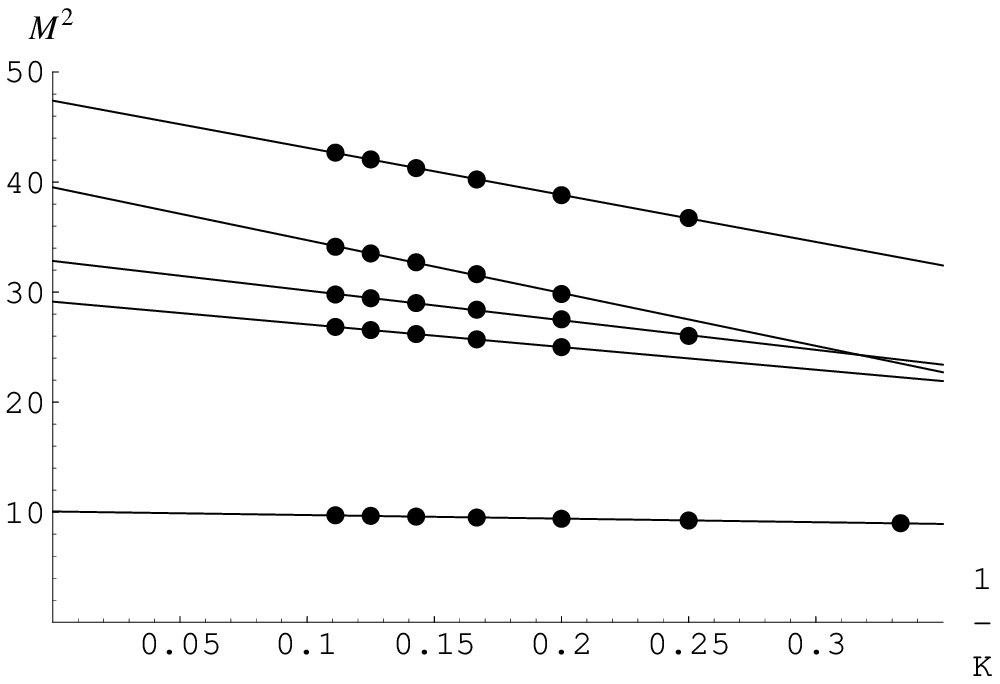,width=7.5cm,angle=0}  \\
(a) & (b)
\end{tabular}
\caption{Low-lying spectrum of the two-dimensional theory 
in units of $g^2 N_c/\pi$ at 
Chern--Simons coupling $h=1.0$ for the (a) $Z_2$ even sector and 
(b) $Z_2$ odd sector.}\label{Fig1}
\end{figure}

The low-energy spectrum, with $h=1.0$, is fit to 
$M^2=M^2_{\infty} +  b(1/K)$.  These fits are shown in Fig.~\ref{Fig1} 
for some of the low-energy states. 
In principle, higher energy states can also be found, but
at these couplings the states just above these are difficult to disentangle
because of level crossings. In the
$Z_2$ even sector we find the continuum masses to be
(in units of $g^2 N_c/\pi$) $M^2_\infty= 4.30$, 18.33, 27.46, and
43.20, whereas in the $Z_2$ odd sector we have
$M^2_\infty= 10.06$, 29.13, 32.83, 39.52, and 47.40.

\begin{figure}
\begin{tabular}{cc}
\psfig{figure=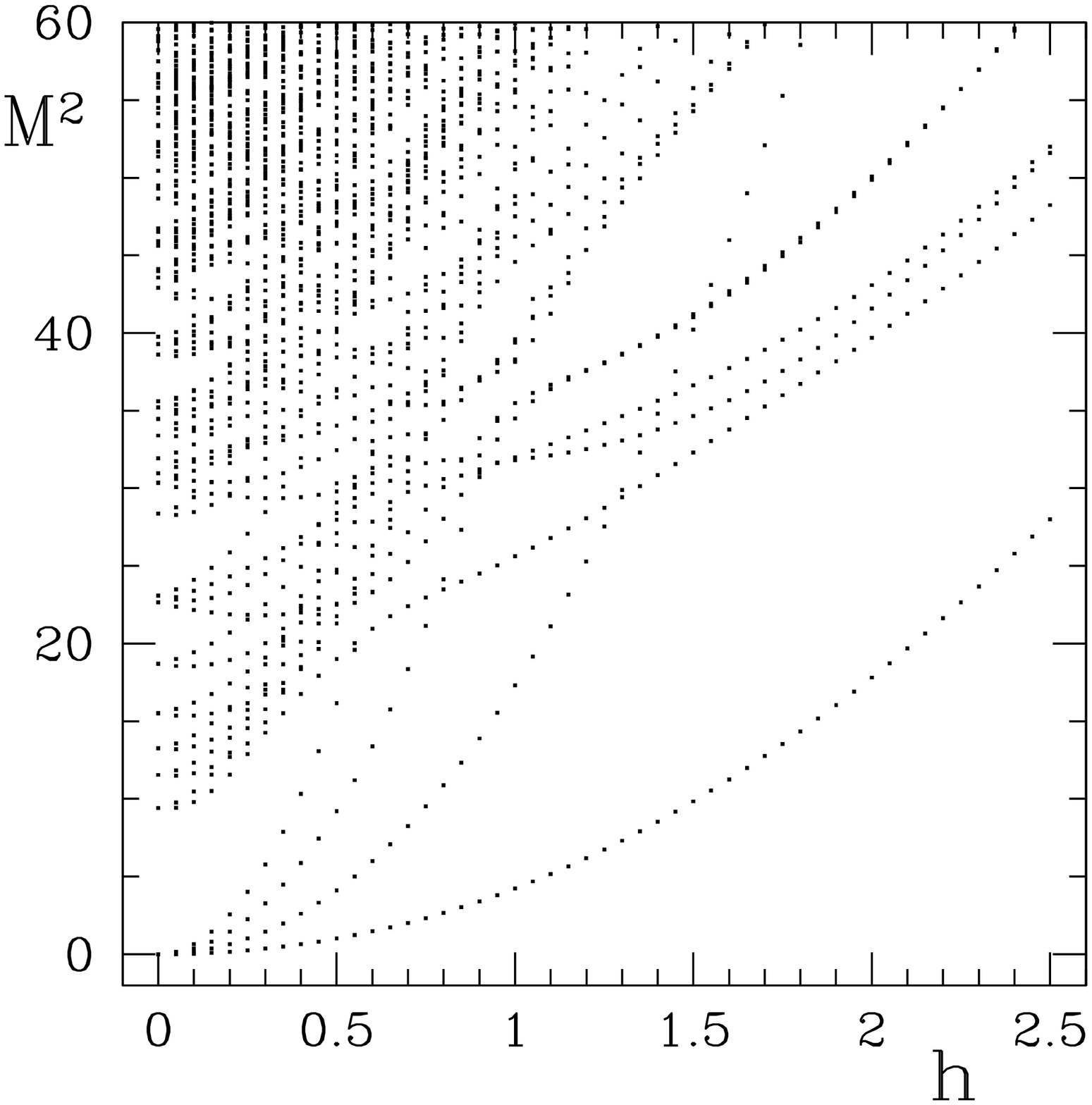,width=7.5cm,angle=0} &
\psfig{figure=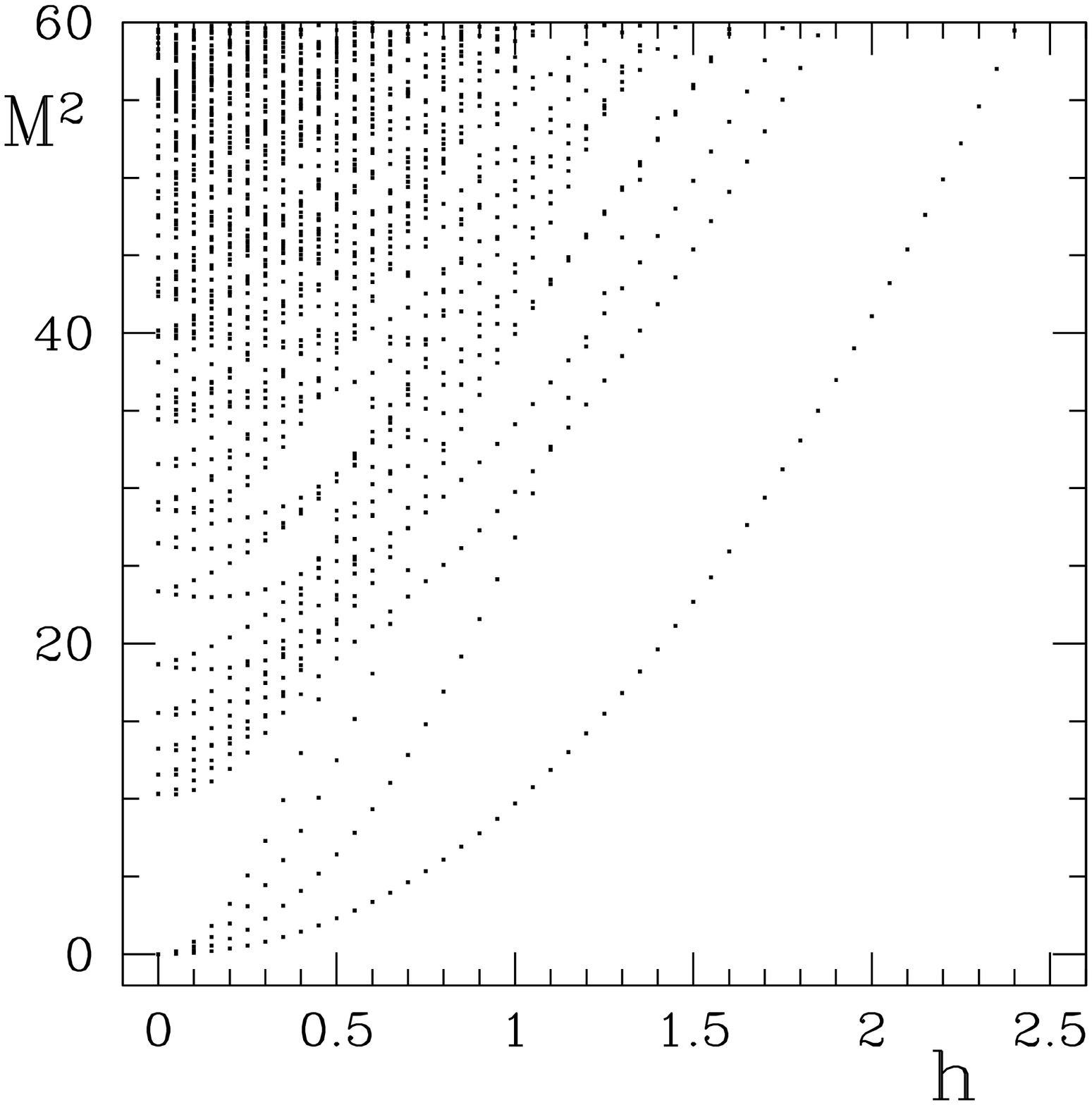,width=7.5cm,angle=0} \\
(a) & (b)
\end{tabular}
\caption{Bosonic spectrum of the two-dimensional theory 
in units of $g^2 N_c/\pi$ at $K=9$
as a function of the Chern--Simons coupling $h$ for the
(a) $Z_2$ even sector and (b) $Z_2$ odd sector.} 
\label{Fig2}
\end{figure}

The CS term in this theory effectively generates a mass proportional to
the CS coupling. Therefore, we expect the low-mass states will
only have a few partons.  This property will be even more apparent as
we increase the CS coupling. This is interesting and important for two
reasons. First, it stands in stark contrast to ${\cal N}=1$ SYM theory,
which is very stringy and has a large number of low-mass states with a large
number of partons. Second, this behavior of the CS theory is
reminiscent of QCD. In Fig.~\ref{Fig2} we plot the spectrum of the theory
as a function of the  scaled CS coupling $h$. We see that the masses of
the bound states grow with the CS coupling, i.e.~the effective mass of the
constituents. We also see that there are a lot of level crossings which
make it hard to follow the trajectories of individual states. 

\begin{figure}
\begin{tabular}{cc}
\psfig{figure=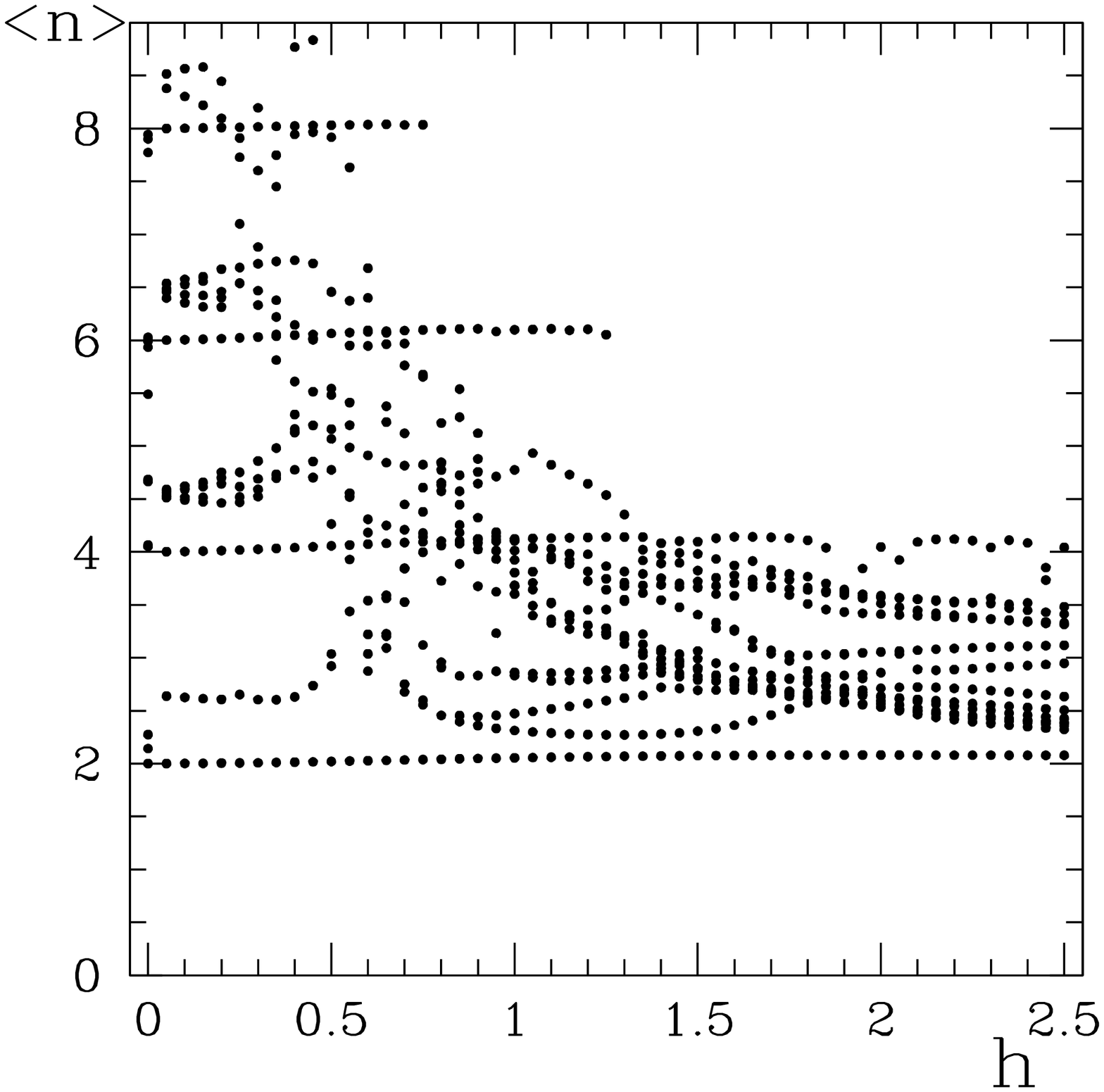,width=7.5cm,angle=0} &
\psfig{figure=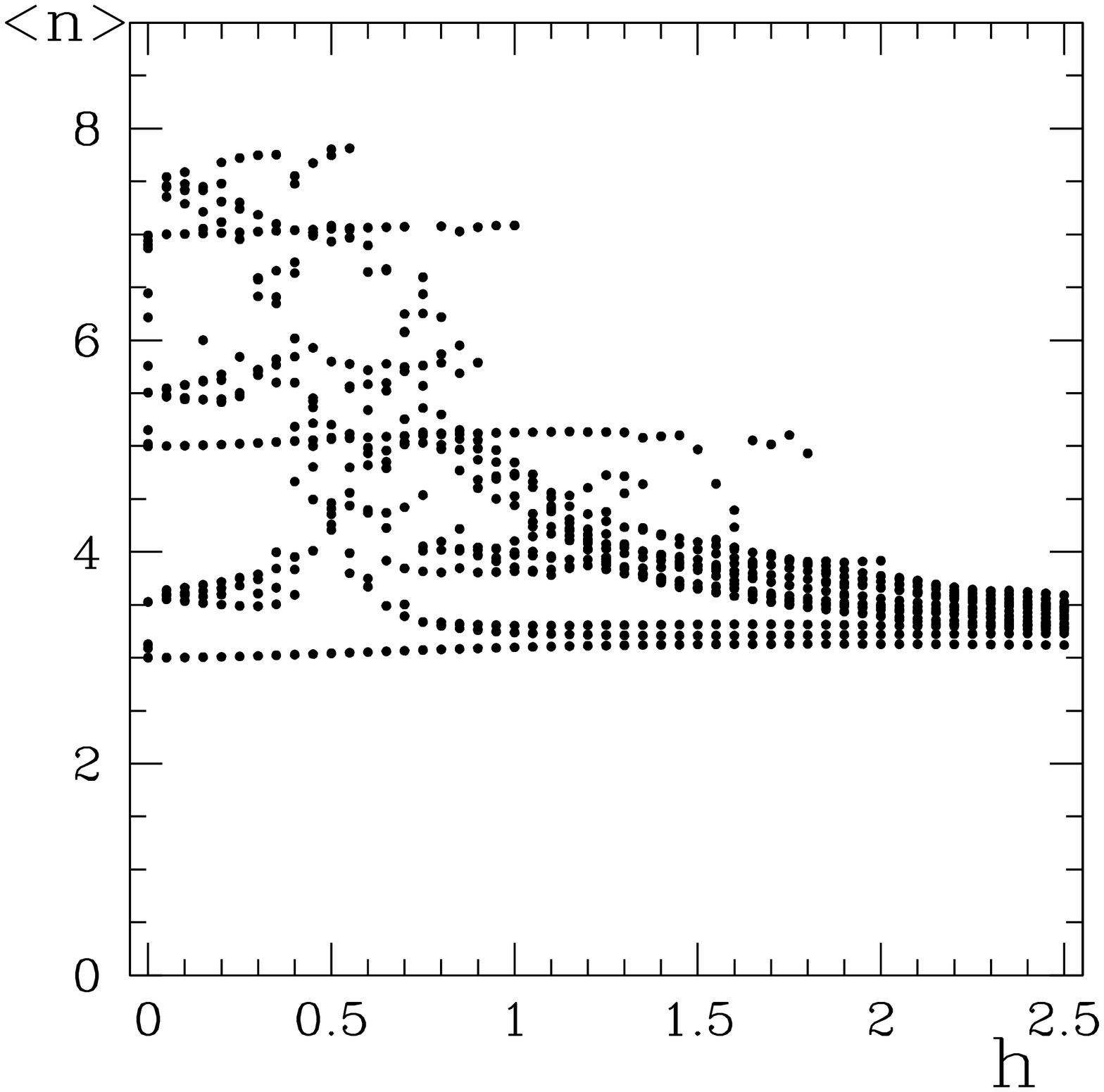,width=7.5cm,angle=0} \\
(a) & (b)
\end{tabular}
\caption{Average parton number of the 15 lowest bosonic states at $K=9$ 
as a function of $h$ for the (a) $Z_2$ even sector and (b) $Z_2$ odd
sector.}
\label{average:n}
\end{figure}

In  Fig.~\ref{average:n} we  plot the average parton number in the 
fifteen lowest states in each sector as a function of the CS coupling for
resolution $K=9$. The average parton numbers range between 2 and the maximally
allowed $\langle n\rangle=9$ at $h=0$ and decrease to below 4
at $h=2.5$. The disrupted trajectories $\langle n\rangle(h)$ are, of course,
due to the level crossings. Most prominently we have crossings at 
$h\approx 0.6$ and $h\approx 1.3$, cf.~Fig.~\ref{Fig2}, which are reflected
in the discontinuities of the $\langle n\rangle(h)$ trajectories at these
points. The apparent lack of states with $\langle n\rangle\approx 2$ 
for $0<h<0.5$ can be explained by the mixing of very light states of
very different parton content. At $h=0$ we have $2(K-1)$ massless states
which have parton numbers all the way up to 9. These states
mix to give the very light states at $h>0$, which eventually are distinct
enough to form independent $\langle n\rangle(h)$ trajectories. 

\begin{figure}
\begin{tabular}{cc}
\psfig{figure=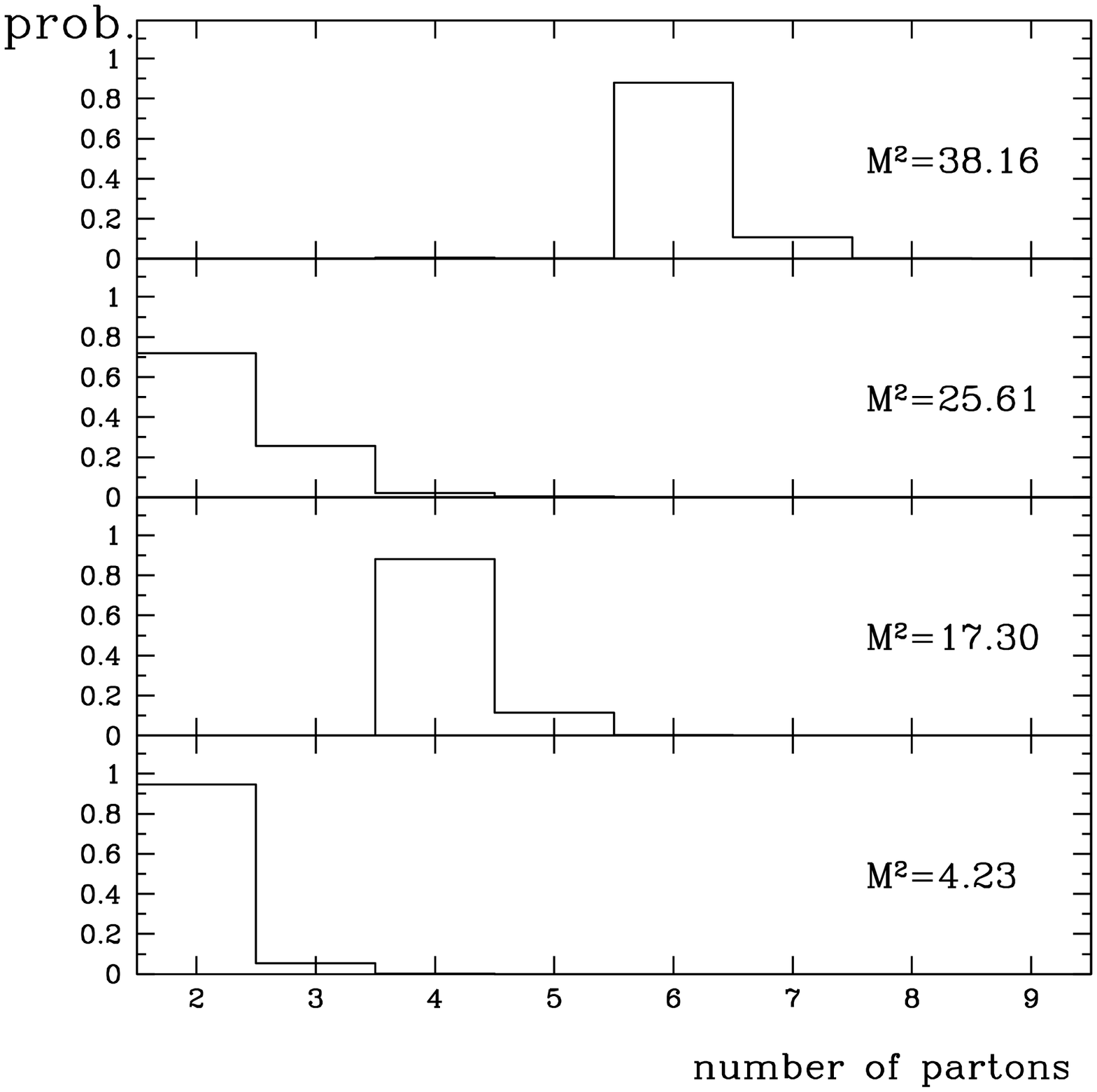,width=7.5cm,angle=0} &
\psfig{figure=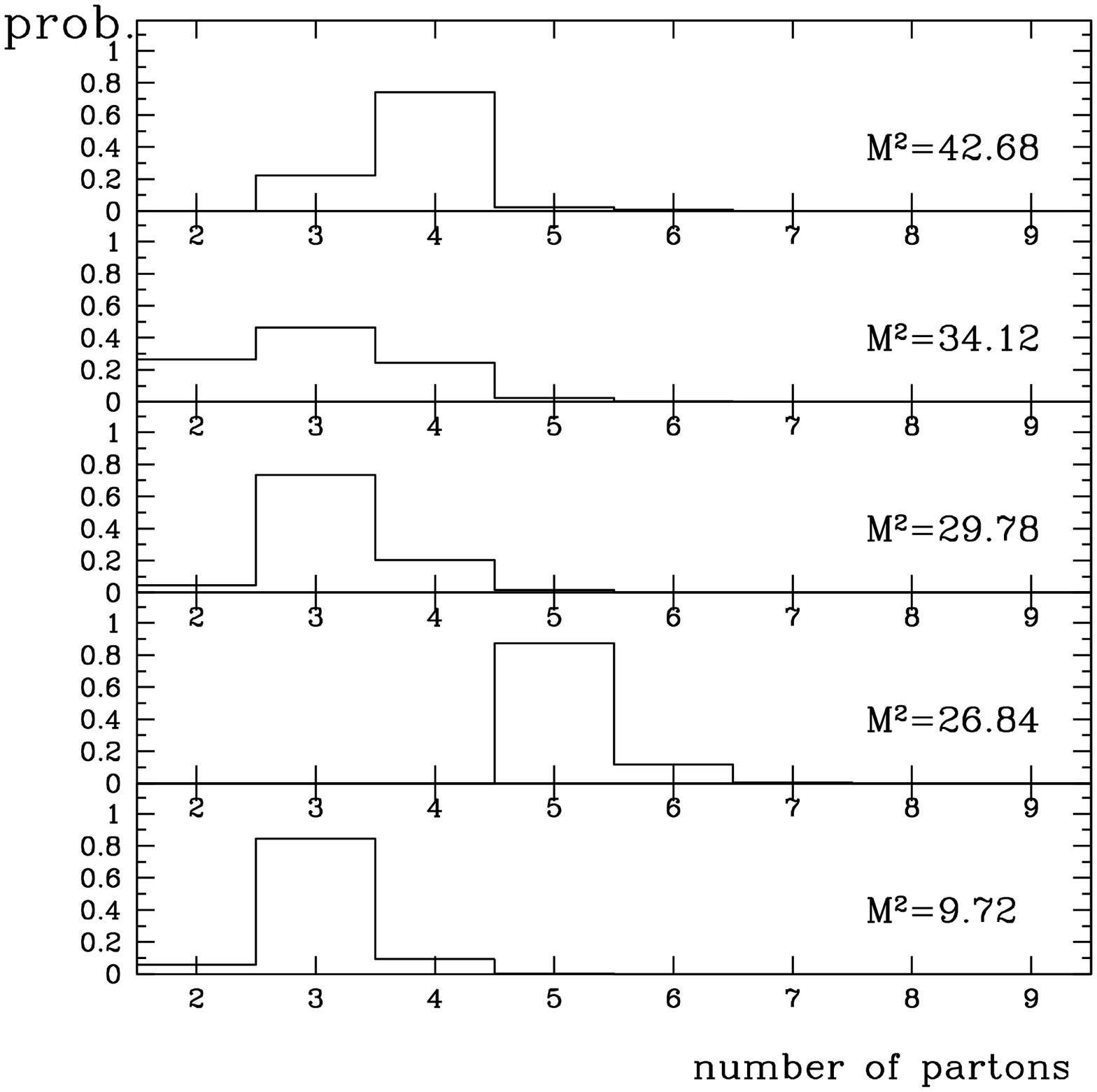,width=7.5cm,angle=0} \\
(a) & (b)
\end{tabular}
\caption{Parton probability distributions for the 9 states in
Fig.~\ref{Fig1} for the (a) $Z_2$ even sector and (b) $Z_2$ odd
sector.}
\label{ppd}
\end{figure}

In Fig.~\ref{ppd} we plot the probability of the nine lowest-energy 
bound states to have a specific number
of partons. In the $Z_2$ even
sector, the two lowest states are nearly pure two-particle and four-particle 
bound states, respectively, while the higher states shown have mixed content. 
We have looked at many of the higher mass states, and we find other nearly
pure states, but almost always with an even number of partons. Similarly in 
the $Z_2$ odd sector we find that the two lowest states are nearly pure
but now have three and five partons. Again the high states shown are of
mixed parton number, but there are other higher mass states not shown that
are nearly pure but almost always have an odd number of partons. The 
probabilities for the degenerate fermionic bound states are not shown, 
but they are virtually identical to these in the respective sectors.  
The fermionic states for the most part just have one of the bosons 
replaced by a fermion.

\begin{figure}
\begin{tabular}{cc}
\psfig{figure=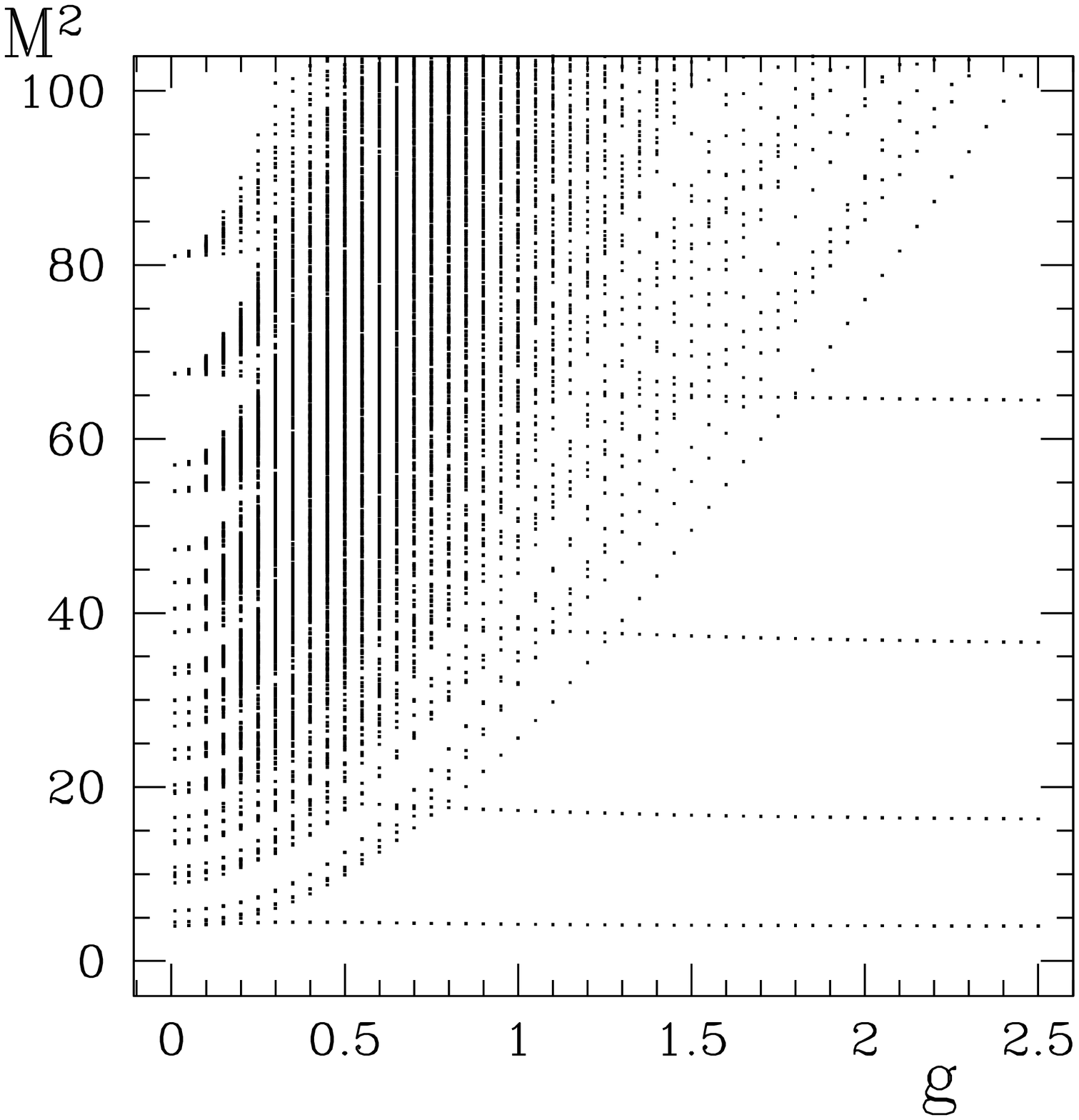,width=7.5cm,angle=0} &
\psfig{figure=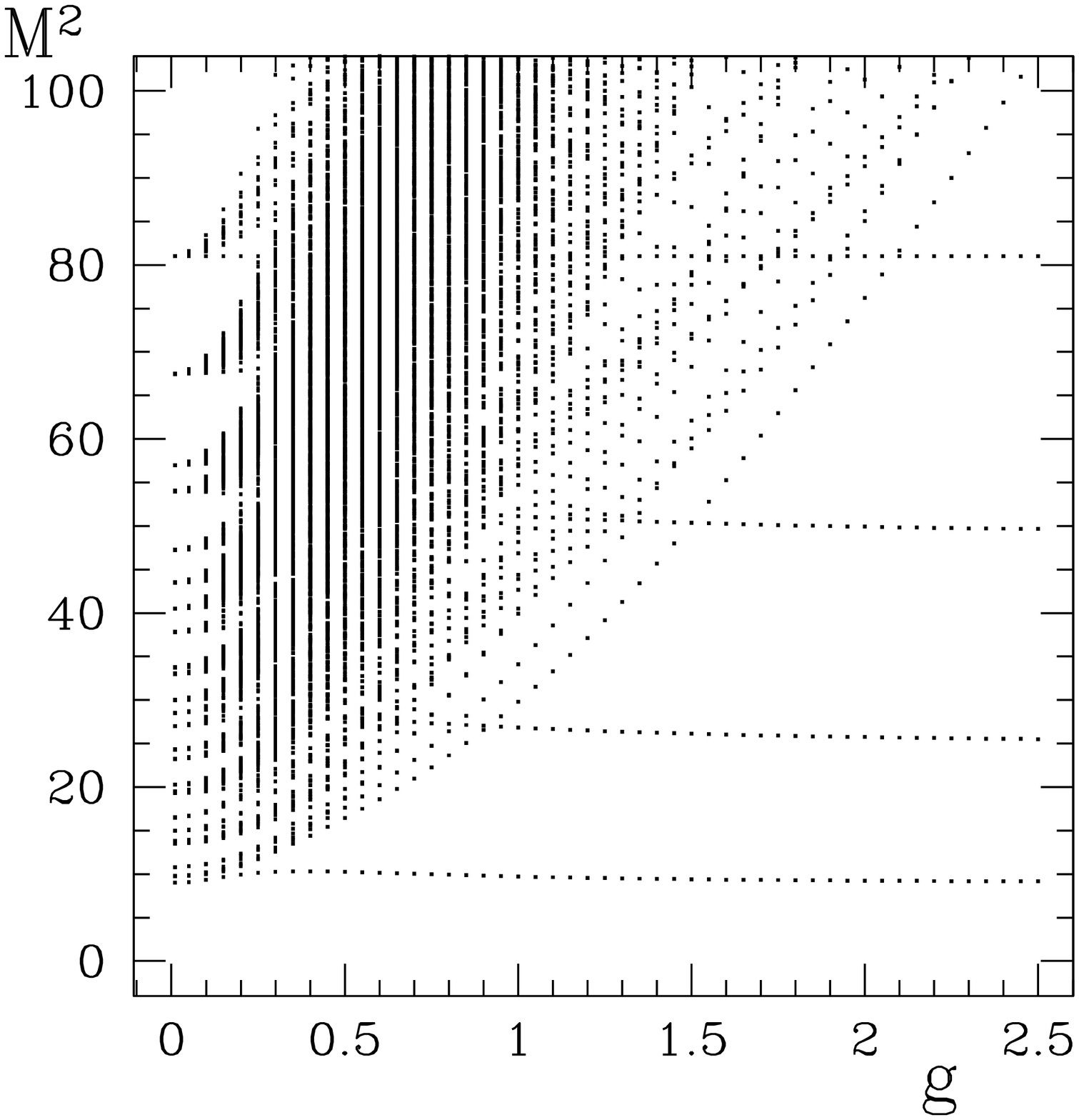,width=7.5cm,angle=0} \\
(a) & (b)
\end{tabular}
\caption{Bosonic spectrum of the two-dimensional theory 
in units of $\kappa^2 N_c$  
as a function of the gauge coupling $g$ at 
fixed Chern--Simons coupling $\kappa$ for the
(a) $Z_2$ even sector and (b) $Z_2$ odd sector.
The values of the gauge coupling are defined in units of 
$\kappa/\sqrt{\pi}$.  The resolution is $K=9$.} 
\label{spec_g0}
\end{figure}

The general structure of the supercharge for this CS theory is
\begin{equation}
Q^- = g Q^-_{\rm SYM}+ \kappa Q^-_{\rm CS}\,,
\end{equation}
where $Q^-_{\rm SYM}$ is the supercharge of the ${\cal N}=1$ SYM theory of 
adjoint fermions and adjoint bosons, which was studied extensively in
\cite{Antonuccio:1998kz}, and $Q^-_{\rm CS}$ is the contribution of the CS
interaction to the supercharge, given in (\ref{qcs}). The Hamiltonian is the
square of the supercharge, and we therefore expect that as a function of 
$g$ and $\kappa$ the spectrum of this theory will grow quadratically in 
both variables. In Fig.~\ref{Fig2} we see this behavior as a function of 
$h=\sqrt{\pi}\kappa/g$ at fixed $g$. In Fig.~\ref{spec_g0} we see
this general behavior at fixed $\kappa$ as a function of $g$ as well. 
There are, however, a
number of very special states visible in this figure. These states are
reflections of the BPS states that we found in the pure SYM
theory \cite{Antonuccio:1998kz}. Since the central charge is zero in that
theory, the BPS states are exactly massless and are annihilated by the
supercharge $Q^-_{\rm SYM}$. The special states that we see in 
Fig.~\ref{spec_g0}
are essentially these BPS states arranged with a fixed number of particles.
Thus the masses of these states are approximately independent of $g$ and 
proportional to the number of partons squared. In this theory at strong 
coupling, clearly the low-mass states are these BPS related states. 
We have already found these BPS states in the (2+1)-dimensional SYM theory
\cite{Haney:2000tk,Antonuccio:1999zu}, and, therefore, we expect that these
states will dominate the low-energy spectrum of the (2+1)-dimensional CS
theory at strong coupling.

We would expect that at $g=0$ all states have masses $M^2=n^2$ 
in units of $\kappa^2 N_c$,
where $n$ is an even(odd) integer in the $Z_2$ even(odd) sector of the 
theory. At $g=0$ only the CS part of the supercharge is present, and it 
is basically a parton number operator. 
What we find, however, are states with well-defined masses,
not only at $M^2=n^2$, but also at intermediate values, which are in general
highly degenerate. The degeneracy is lifted as the coupling $g$ is turned on,
and in the strong coupling limit all but the BPS states decouple, 
as can be seen in Fig.~\ref{spec_g0}. 
A closer inspection of the convergence of the eigenvalues
immediately reveals that the states at $g=0$ are actually multi-particle 
states built out of constituents with ``mass'' proportional to $\kappa$. 
This is very much like what happens in the large $N_f$ limit of adjoint 
QCD$_2$ \cite{UT}. Actually, the DLCQ spectra $M^2_i(K)$,
as a function of resolution $K$, look almost exactly like the 
analogous plots in Ref.~\cite{UT}, and we suppress them here.
They are consistent with the DLCQ expression for the kinetic energy of 
free particles. For example, the two-particle formula is
\begin{equation}
\frac{M^2(K)}{K}=\frac{M^2(n)}{n}+\frac{M^2(K-n)}{K-n}\,.
\end{equation}
Of course, the physics of the two theories is 
very different, and this easily explains 
the few differences of the spectra. The crucial point is that we thus
{\em completely} understand the spectrum at $g=0$ --- and also at large $g$.
It is very interesting that we produce all 
multi-particle states at $g=0$, and that only the mass of the 
lowest state in each set of
states with fixed parton number stays finite in the large-coupling limit. 
In this sense the theory is dominated by the underlying supersymmetry,
although the BPS symmetry seems slightly broken.

In Table~\ref{table1} we show the masses and the average numbers of
partons and fermions in several of the lowest mass states at resolution
$K=9$ for both the bosonic and the fermionic sectors of the theory
as well as the $Z_2$ even and odd sectors.
While masses and average parton numbers are identical between
bosonic and fermionic sectors,
the average fermion numbers are obviously different.
\begin{table}[ht]
\caption{Average parton and fermion numbers for the lowest 15 states 
in the $Z_2$ even and odd sectors at $K=9$.
While masses and average parton numbers are identical in the bosonic
and fermionic sectors,
the average fermion numbers $\langle n_F\rangle_B$ and 
$\langle n_F\rangle_F$, respectively, are obviously different.
\label{table1}}
\begin{tabular}{r|cccc|cccc}
No.& $M^2_{Z_2=+1}$ & $\langle n\rangle$ &  $\langle n_F\rangle_B$ &
$\langle n_F\rangle_F$ &  
$M^2_{Z_2=-1}$ & $\langle n\rangle$ &  $\langle n_F\rangle_B$ & 
$\langle n_F\rangle_F$ \\ 
\hline 
1  &  4.2344 & 2.0543 &  0.1095 & 1.1061 &  9.7197 & 3.0982 &  0.2012& 1.1945\\
2  & 17.3049 & 4.1229 &  0.2539 & 1.2449 & 26.8372 & 5.1282 & 0.2674 & 1.2572\\
3  & 25.6130 & 2.3138 &  1.7763 & 1.3675 & 29.7822 & 3.2386 &  1.8603& 1.2995\\
4 & 31.8181 & 4.1052 &  2.0004 & 1.4683 & 34.1192 & 3.3053 &  0.5905 &1.1407 \\
5 & 32.0114 & 2.4724 &  0.7135 & 1.1135  & 39.9376 & 4.5267 &  2.0000&1.3423\\
6 & 34.4887 & 3.9241 &  1.9292 & 1.3323  & 40.5369 & 4.7211 &  0.6094&1.2480\\
7 & 35.4751 & 3.8048 &  0.6713 & 1.1965  & 42.6751 & 3.8168 &  1.9404&2.5317 \\
8 & 38.1566 & 6.0990 &  0.2467 & 1.2336  & 43.8869 & 4.8453 &  0.5371&1.2022 \\
9 & 38.3108 & 3.6778 &  1.9026 & 1.3923  & 44.4106 & 4.7409 &  1.9689&1.3118\\
10 & 39.3822 & 4.0123 &  0.6590 & 1.2340 & 46.8487 & 3.9974 &  1.9260&1.3716\\
11 & 39.5958 & 2.8320 &  1.8687 & 2.3274 & 48.1737 & 3.9713 &  1.8218&1.2551\\
12 & 44.5659 & 3.6840 &  1.8453 & 1.3832  & 49.2253 & 4.1032 &  0.5985&1.1508\\
13 & 46.2064 & 3.6026 &  0.7155 & 1.1640& 49.7721 & 4.4400 &  2.1560&1.6061\\
14 & 47.5226 & 2.8625 &  1.6782 & 1.2451 &  51.1264 & 7.0836 &  0.1898&1.1802\\
15 & 48.3041 & 4.7750 &  1.9301 & 2.6179 &  51.9753 & 3.8548 & 2.0888&1.5990\\
\end{tabular}
\end{table}

\begin{figure}
\centerline{
\psfig{figure=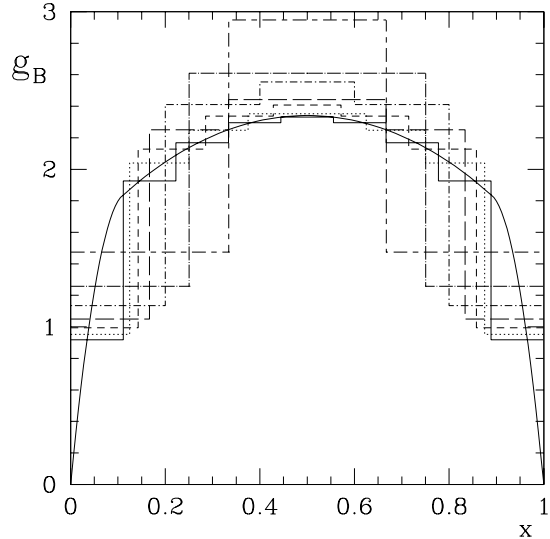,width=7.5cm,angle=0}
}
\caption{Convergence of the structure function of the lightest 
state of the theory as the resolution $K$ is increased from 3 to 9. 
The smooth solid line is a spline interpolation to the
data for $K=9$ and the conjectured points at $x=0$ and $x=1$.
The scaled Chern--Simons coupling $h$ is equal to 1.}
\label{convergenceSF}
\end{figure}

\begin{figure}
\begin{tabular}{cc}
\psfig{figure=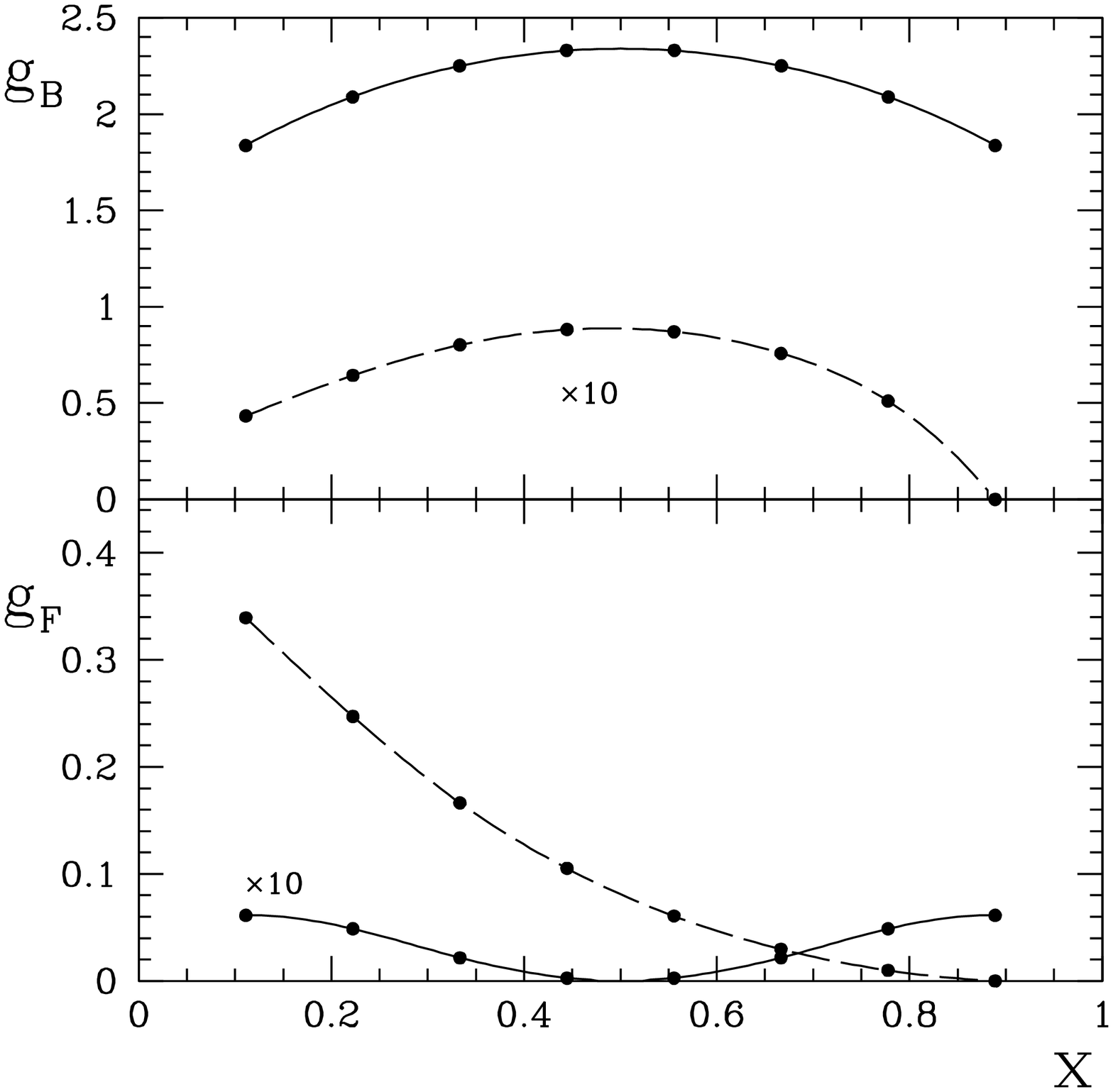,width=7.5cm,angle=0} &
\psfig{figure=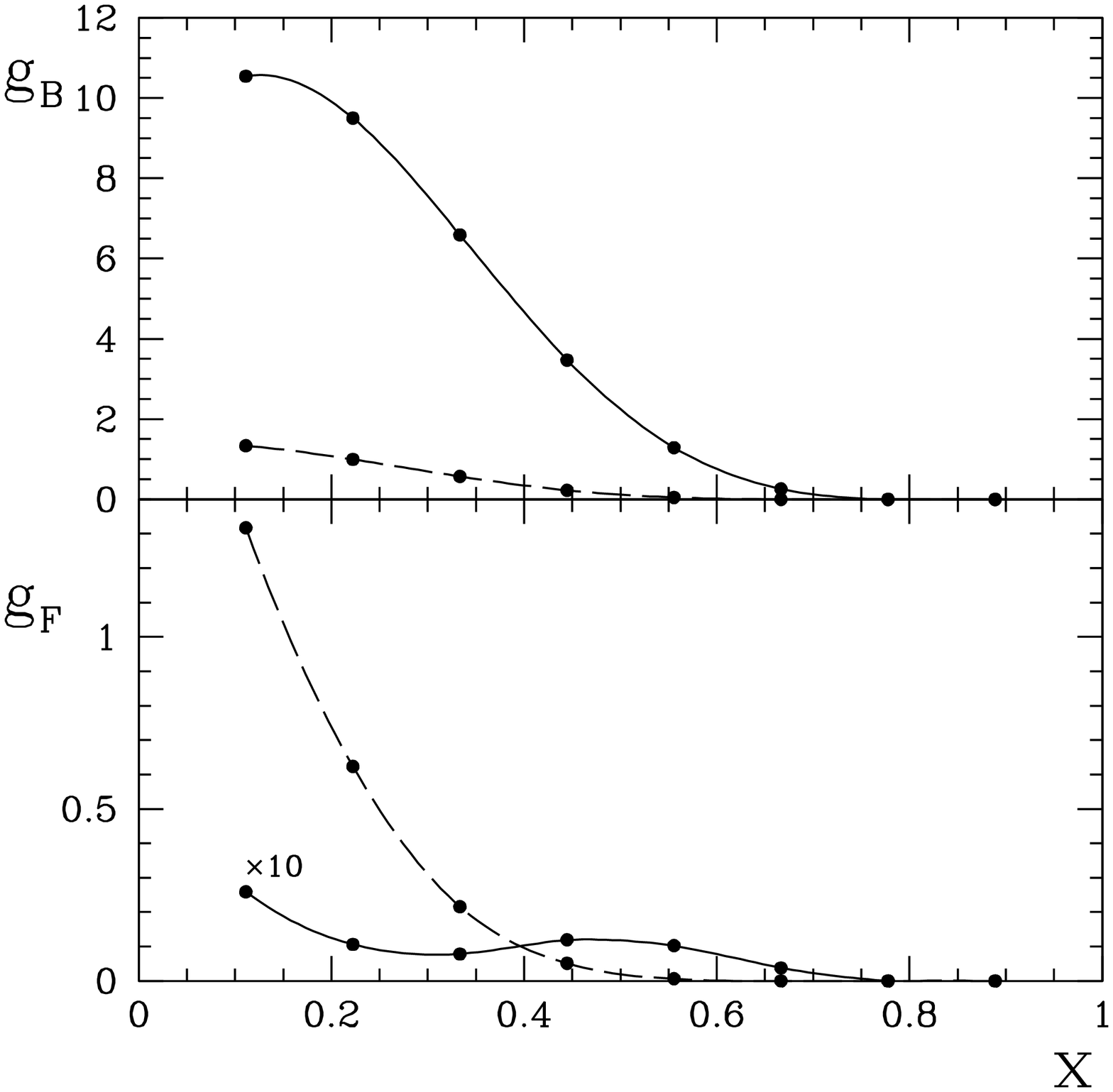,width=7.5cm,angle=0} \\
(a) & (b) \\
\psfig{figure=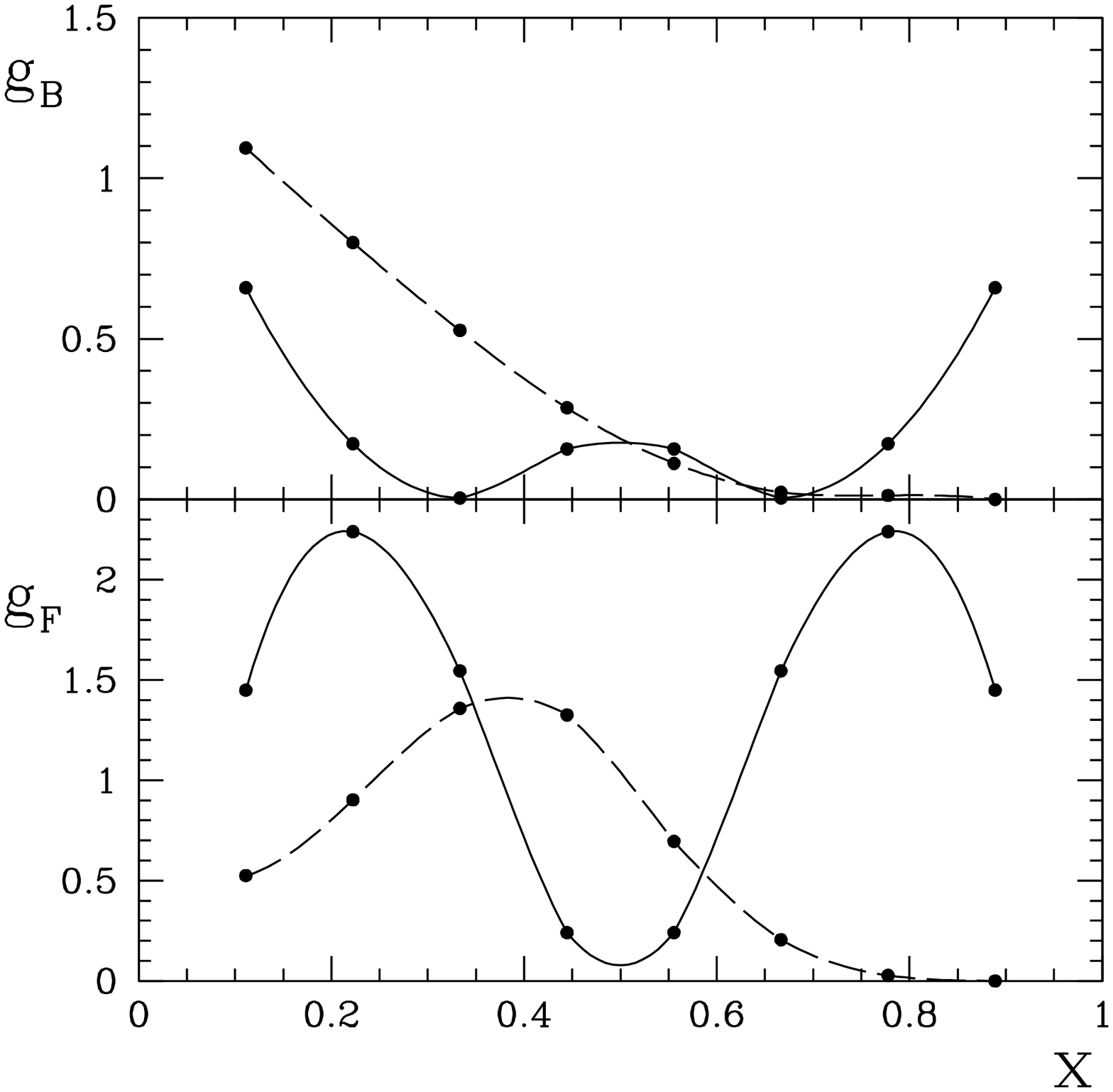,width=7.5cm,angle=0} &
\psfig{figure=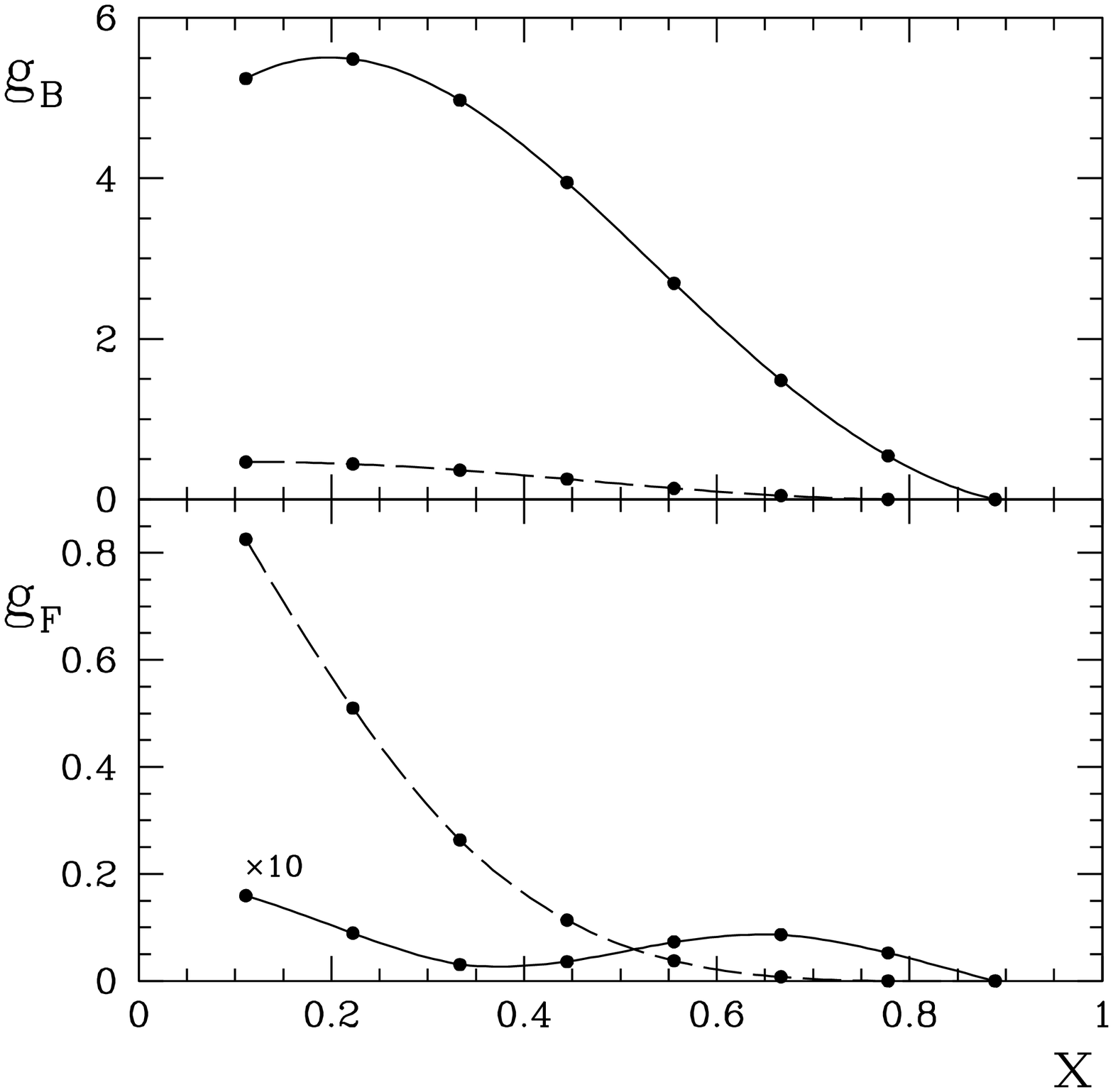,width=7.5cm,angle=0} \\
(c) & (d)
\end{tabular}
\caption{Structure functions of the lightest three $Z_2$ even
{\em bosonic} states, with (a)  $M^2=4.23$, (b) $M^2=17.30$, (c) $M^2=25.60$,
and (d) the lightest $Z_2$ odd state with $M^2=9.72$.  These masses
are in units of $g^2N_c/\pi$ at fixed $h=1$, and the numerical 
resolution is $K=9$. The top half of each plot shows
the bosonic structure functions, and the bottom half,
the fermionic structure functions. 
Dashed lines represent sea structure functions; solid lines represent 
structure functions in the two-parton sector, for (a) and (c);
in the four-parton sector for (b); 
and in the three-parton sector for (d).
For the purpose of visibility, the following functions have
been multiplied by 10: In (a), the fermionic two-parton and bosonic
sea structure functions; in (b), the fermionic four-parton structure
function; and in (d), the fermionic three-parton structure function.
}\label{sfs}
\end{figure}

\begin{figure}
\begin{tabular}{cc}
\psfig{figure=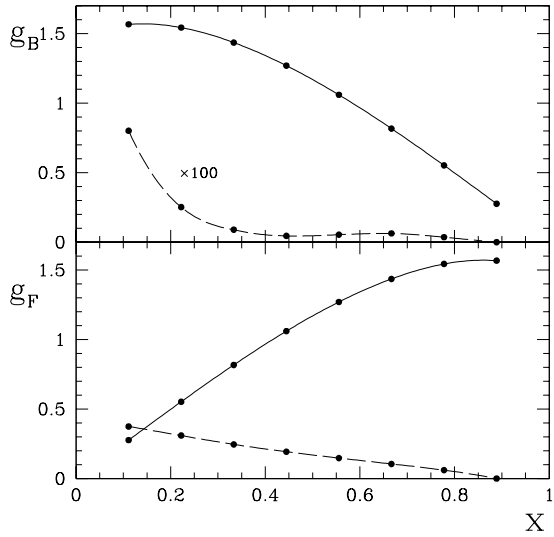,width=7.5cm,angle=0} &
\psfig{figure=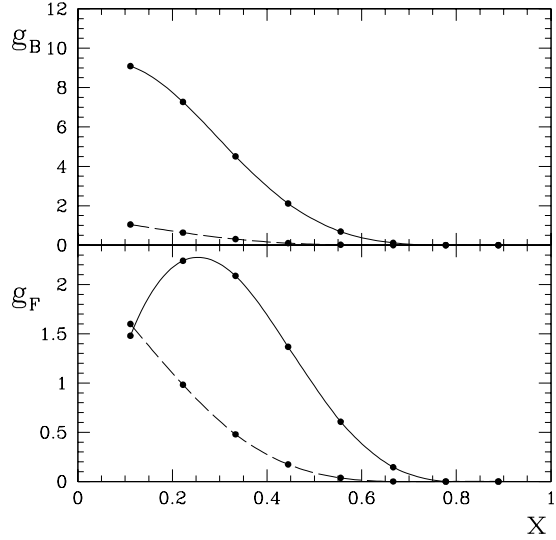,width=7.5cm,angle=0} \\
(a) & (b) \\
\psfig{figure=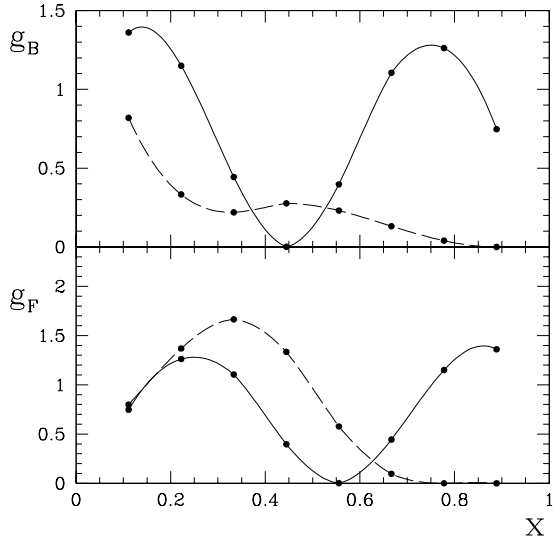,width=7.5cm,angle=0} &
\psfig{figure=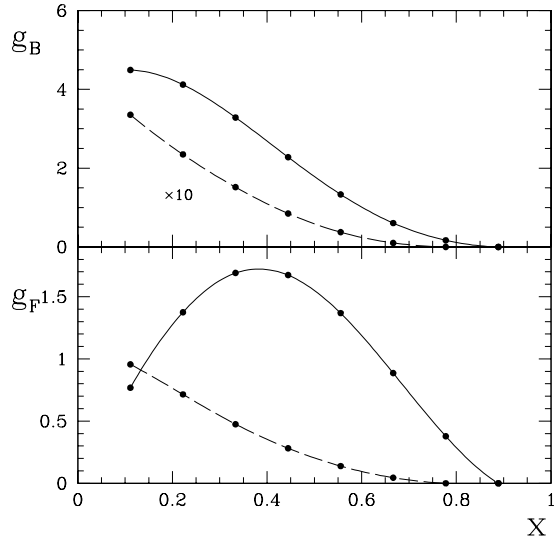,width=7.5cm,angle=0} \\
(c) & (d)
\end{tabular}
\caption{Same as Fig.~\ref{sfs} but for fermionic states,
with only the bosonic sea structure functions in (a) and (d)
rescaled by 100 and 10, respectively.
}\label{fermionicSFs}
\end{figure}

It is very instructive
to look at the structure functions for the computed bound states,
particularly since they are very QCD-like. We use a standard definition of
the structure functions
\begin{eqnarray}
\hat{g}_A(x)&=&\sum_q\int_0^1 dx_1\cdots dx_q 
\delta\left(\sum_{i=1}^q x_i-1\right)
\sum_{l=1}^q \delta(x_l-x)\delta^A_{A_l}
|\psi(x_1,\ldots x_q)|^2\,.
\end{eqnarray}
Here $A$ stands for either a boson or a fermion.
The sum runs over all parton numbers $q$, and the Kronecker delta 
$\delta^A_{A_l}$ selects partons with matching statistics $A_l$.
The discrete approximation $g_A$ to the structure function $\hat{g}_A$
with harmonic resolutions $K$ is
\begin{eqnarray}
{g}_A(n)&=&\sum_{q=2}^K\sum_{n_1,\ldots,n_q=1}^{K-q}
\delta\left(\sum_{i=1}^q n_i-K\right)
\sum_{l=1}^q \delta^{n_l}_n\delta^A_{A_l}
|\psi(n_1,\ldots n_q)|^2\,.
\end{eqnarray}
The functions $g_{A}(n)$ are normalized so that summation over the 
argument gives the average boson or fermion number; 
their sum is then the average parton 
number, and we compute these sums as a test. We scale the momentum
distribution to the total longitudinal momentum and plot the structure
functions as functions of the longitudinal momentum fraction
$x=k^+/P^+$ carried by an individual parton.  Several
structure functions are shown in Figs.~\ref{convergenceSF},
\ref{sfs}, and \ref{fermionicSFs}.

We analyze the 
structure functions of the lowest bound states of this
theory in some detail. There are, after all, very few first-principles 
calculations of structure functions for gauge theories.
We are particularly interested in the states that have a dominant  number
and type of particle, which is what one generally refers to as  the
valence partons. We will look at both the valence and sea structure
functions. Unfortunately, we cannot go to high enough resolution here to 
get the wee parton structure functions at small $x$. This is not a
limitation of our computing power, but rather of our code which is
designed for 2+1 dimensional problems. We estimate that a rewritten code
could go to resolution 50 with our present computing power.

We have seen that the eigenvalues of the theory converge rapidly, and we 
therefore expect that the same will be true for the wave functions. We can
demonstrate this convergence as follows.  We consider the structure function 
of the lowest mass state in the $Z_2$ even sector at various values of the
resolution $K$ and focus on the valence boson distribution of the lowest 
boson state. This state is dominantly a two-gluon bound state. For this
demonstration we assume that the distributions vanish at $x=0$ and
$x=1$, which  seems sensible for valence partons.  We see in
Fig.~\ref{convergenceSF} that  the structure function appears to converge
to a well-defined curve, supporting the notion that the wave function as
well as the eigenvalues converge rapidly. 

This structure function for the valence gluons (which is also shown 
in Fig.~\ref{sfs}(a)) in the lowest bosonic state is 
peaked at $x=0.5$, as one would expect for a two-gluon bound state and
interestingly the distribution is quite broad. As we know,  in QCD a
glueball state will naturally mix with the fermions in the theory.  In
Fig.~\ref{sfs}(a) we see the sea fermion structure function for this state
as well as the sea boson structure functions. We find that the fermion
distributions tend to peak at low $x$ and are relatively small as
compared to the valence distributions. The boson sea distribution, which
is primarily from the three-parton component of the wave function, does not peak
at small $x$. The reason is that the small-$x$ component of the three-parton
wave function has a nearly equal mixture of fermion pairs and boson pairs
with small $x$. These small-$x$ fermions are seen in the fermion distribution.

The structure functions of the second lowest bound states in the $Z_2$ even
bosonic sector are shown in Fig.~\ref{sfs}(b). This is a four-gluon
bound state, the analog of what is referred to as an ``odd ball" in QCD. The
valence-gluon structure function is expected to peak at about $x=0.25$,
but we see that the actual peak is at smaller values of $x$. The various 
sea distributions are shown in Fig.~\ref{sfs} and again appear to peak at 
small $x$. 

The third state in the $Z_2$ even bosonic sector is a highly mixed state
containing a nearly equal mixture of fermions and gluons and does not 
appear to have the simple valence structure of the lowest two states. 
In Fig.~\ref{sfs}(c) we show separately the distribution of the fermions 
and bosons. It is not surprising that we find such a mixed state, since 
in the supersymmetric theory the fermion and boson are treated on an equal footing, but rather it {\em is} surprising that the lowest states have a 
valence structure. 

It is also interesting to discuss the fermionic states. 
At first sight it might be surprising that their structure functions,
shown in Fig.~\ref{fermionicSFs}, are so different from those of
their bosonic partners. Even more so since they have the same mass 
and average parton number, guaranteed by supersymmetry. 
A closer look reveals, however, that the new structures are due to 
the different combinatorics induced by the ``extra'' fermionic parton.
For example, the valence structure functions of the state with $M^2=25.61$
are symmetric around $x=0.5$ 
in the bosonic sector, and almost symmetric in the fermionic sector. 
The asymmetry is clearly induced by the need to have an additional 
parton for the statistics. 
For the states with $M^2=9.72$ and $M^2=17.30$ it seems as though the 
curves stay more or less the same, except that the fermion valence 
structure function is greatly enhanced.
The lightest state has an interesting valence structure function:
It is exactly as probable to find a boson with momentum fraction $x$,
as it is to find a fermion with $1-x$.  

\section{Summary} \label{sec:Summary}

We have presented an analysis of dimensionally reduced supersymmetric 
YM theory with a CS term using the numerical method SDLCQ, which exactly 
preserves the supersymmetry of this theory. We constructed finite dimensional
representations of the superalgebra and from them a finite
dimensional Hamiltonian which we solved numerically. As we go to higher
dimensional representations, the solutions converge rapidly. 

>From these solutions we extracted the properties of the bound states of 
this theory.  We found that the bound states are very different from 
the bound states of ${\cal N}=1$ SYM in 1+1 dimensions. The bound states of 
SYM are characterized by their stringy nature, that is a set of states where 
the states with the most partons have the lowest energy.  
The CS theory is generally very QCD-like.  The states with
more partons tend to be more massive, and many of the low-mass states
have a valence-like structure. These states have a dominant component of
the wave function with  a particular number and type of parton. We have
found the spectrum of these states and studied it as a function of the CS
coupling.  We found that, as expected, the CS coupling behaves as an
effective mass.  As it increases we see that the masses of the bound
states generally increase, and the average number of partons of the
low-mass bound states tends to decrease. 

Also, the SYM theory has an interesting set of massless BPS states. 
These states are reflected in the CS theory as a set of states whose
masses are approximately independent of $g$ and equal to the square of the
sum of the CS masses of the partons in the bound state. 
Since these BPS states are also present in the
(2+1)-dimensional SYM theory, we expect to see their reflection 
in the (2+1)-dimensional CS theory. 

We have investigated the structure functions for these bound states and 
found that they behave in very interesting ways. The distribution for the
states with a valence structure are peaked near the inverse of the number
of valence partons, and the sea distribution appears to peak at small $x$ in
most if not all cases. We also see strongly  mixed states with interesting
double-humped structure functions. There  have been interesting
conjectures about structure functions of this type in QCD \cite{cz}, and
it is interesting that we actually find such a structure function in the
solution of an actual gauge theory. 

In summary, the SDLCQ solutions of dimensionally reduced supersymmetric
CS theory are very interesting, maybe more interesting, at least with
respect to their QCD-like structure, than are the solutions for pure 
supersymmetric SYM theory. Clearly this provides a strong base and 
encouragement to move on to 2+1 dimensions.

\acknowledgments
This work was supported in part by the U.S. Department of Energy.
 

\end{document}